\pdfoutput=1

\documentclass[11pt,twoside,a4paper,cmspaper,final,collab]{cms-tdr}
\def\svnVersion{175876}\def\cmsCernNo{2012-225}\def\cmsCernDate{\today}\def\cmsMessage{Submitted to Physics Letters B}

\begin{document}\cmsNoteHeader{EXO-11-060}

\hyphenation{had-ron-i-za-tion}
\hyphenation{cal-or-i-me-ter}
\hyphenation{de-vices}

\RCS$Revision: 153129 $
\RCS$HeadURL: svn+ssh://svn.cern.ch/reps/tdr2/papers/EXO-11-060/trunk/EXO-11-060.tex $
\RCS$Id: EXO-11-060.tex 153129 2012-10-16 14:40:01Z duggan $
\newlength\cmsFigWidth
\setlength\cmsFigWidth{0.95\columnwidth}
\cmsNoteHeader{EXO-11-060} 
\title{Search for three-jet resonances in pp collisions at $\sqrt{s}=7\TeV$}

\date{\today}

\abstract{
    Results are reported from a search for the production of three-jet resonances in $\Pp\Pp$
    collisions at a center-of-mass energy $\sqrt{s}=7\TeV$. The study uses the
    data sample collected by the CMS experiment at the LHC in 2011, corresponding to an
    integrated luminosity of 5.0\fbinv. Events with high jet multiplicity and a large
    scalar sum of jet transverse momenta are analyzed for the presence of resonances in
    the three-jet invariant mass spectrum. No evidence for a narrow resonance is found in
    the data, and limits are set on the
    cross section for gluino pair production in an R-parity-violating supersymmetry model,
    for gluino masses greater than 280\GeV. Assuming a
    branching fraction for gluino decay into three jets of 100\%,
    gluino masses below 460\GeV are excluded at 95\% confidence level. These
    results significantly extend the range of previous limits.
}

\hypersetup{%
pdfauthor={CMS Collaboration},%
pdftitle={Search for three-jet resonances in pp collisions at sqrt(s) = 7 TeV},%
pdfsubject={CMS},%
pdfkeywords={CMS, physics, three-jet resonances, R-parity violation}}

\maketitle 

Many extensions of the standard model (SM) predict the existence of new,
strongly interacting particles that lead to final states with high jet
multiplicities. Examples of such particles include gluinos in
$R$-parity-violating (RPV) SUSY models~\cite{sixjets,multijets}, or
more generally any colored fermion resonance that transforms as an octet
under $SU(3)_\textrm{C}$~\cite{tcgut,exoquarks,unichiral,altchiral}.
Despite large production cross sections predicted by many of these
models, the vast majority of current experimental searches are
insensitive to them because they use signal selection criteria based
on low jet multiplicities, such as dijet resonances, or require the
presence of large missing transverse energy.
While this strategy has proved effective in establishing cross section
limits for many models of physics beyond the SM up to TeV mass
scales~\cite{susy,dijets}, the currently published limits for
models predicting high jet multiplicity
final states~\cite{Ruderman:2012jd,Kumar:2011tj,Kats:2012ym}
are much less stringent.
Specifically, there are only two published results from
searches involving six-jet final states at hadron colliders,
one from the Tevatron~\cite{cdfmultijets} and the other from the
Large Hadron Collider (LHC)~\cite{cmsmultijets}, which exclude RPV
gluinos with masses below 144\GeV and between 200 and 280\GeV,
respectively.

This Letter presents the results of a search for three-jet
resonances in multijet events in proton-proton ($\Pp\Pp$) collisions
at $\sqrt{s}=7\TeV$. The event sample was collected using
the Compact Muon Solenoid (CMS) detector~\cite{cmsdet} at the LHC.
The integrated luminosity is $4.98\pm0.11\fbinv$~\cite{lumi},
corresponding to the CMS data sample recorded in 2011. In this
extension of previous CMS results~\cite{cmsmultijets} using
35\pbinv of data, events with at least six
high-transverse-momentum ($\pt$) jets are investigated
for evidence of three-jet resonances using the jet-ensemble
technique~\cite{cdfmultijets, cmsmultijets}.

The CMS detector is a multi-purpose apparatus described in detail in
Ref.~\cite{cmsdet}. A high-resolution silicon pixel and
strip tracker, immersed in the 3.8\unit{T} magnetic field of the
superconducting solenoid, provides charged particle tracking coverage for
$|\eta|<2.5$, where $\eta=-\ln[\tan(\theta/2)]$ is the
pseudorapidity, and $\theta$ is the polar angle measured with
respect to the counterclockwise proton beam direction.
Jet energy deposits are measured using electromagnetic (ECAL)
and hadronic (HCAL) calorimeters. The finely-segmented lead-tungstate
crystal ECAL consists of a barrel and two endcap regions.
The ECAL barrel covers the pseudorapidity range $|\eta| < 1.4$ with a granularity of
$\Delta\eta \times \Delta\phi = 0.0174 \times 0.0174$, where $\phi$ is the
azimuthal angle, measured in radians, and the endcaps
cover $1.4 < |\eta| < 3.0$.
The HCAL extends out to an $|\eta|$ of about 5.0. Its central and endcap regions
consist of brass/scintillator
sampling calorimeters that cover $|\eta| < 3.0$ with a granularity of
$\Delta\eta \times \Delta\phi = 0.087 \times 0.087$ for central rapidities.
Further coverage is provided by a steel/quartz-fiber Cherenkov calorimeter.
Muons are measured in gas detectors embedded in the steel return yoke
outside the solenoid.

Events are recorded using a two-tiered trigger system~\cite{cmsdet}.
Objects satisfying the requirements at the first level
are passed to the high-level trigger (HLT) which further
reduces the total recorded rate. Triggers based
on the scalar sum of all transverse energy from jets ($\HT$),
reconstructed using only calorimeter information, are
used to select events. To reduce the effects of multiple $\Pp\Pp$
interactions in the same bunch crossings (pileup), which produce
a large number of low-$\pt$ particles, jets considered in the
HLT selection are required to have $\pt >40$\GeV. The HLT
threshold for $\HT$ ranged from 350 to 750\GeV
over the course of the 2011 data collection period,
depending on the instantaneous luminosity.

A further set of criteria is imposed on events passing the
trigger. Events are required to contain at least one
well-reconstructed primary event vertex~\cite{goodPV}. Jets are
required to have $\pt > 70\GeV$ and $|\eta| < 3.0$,
which suppresses background and minimizes pileup effects.
Events with pair-produced gluinos are expected to have high
jet multiplicity and large values of $\HT$. We
therefore require events to contain at least six jets, and
to have a total scalar sum of jet $\pt$ exceeding 900\GeV.
The latter requirement ensures that the trigger is fully
efficient for these events.

Individual objects reconstructed using the CMS particle-flow
algorithm~\cite{pflow} serve as input for jet reconstruction.
The particle-flow algorithm~\cite{pflow} combines calorimeter
information with reconstructed tracks to identify
individual particles such as photons, leptons, and both neutral
and charged hadrons within jets. The energy of photons is obtained
directly from the calibrated ECAL measurement. The determination
of the energy of electrons comes
from a combination of the track momentum at the main
interaction vertex, the corresponding ECAL cluster energy, and the
sum of all energy from bremsstrahlung photons associated to the track. The
energy of muons is obtained from the corresponding track momentum.
Charged hadron energy is determined from a combination of the track
momentum and the corresponding ECAL and HCAL energies, corrected for
hardware zero-suppression effects, and calibrated for the non-linear
response of the HCAL. Finally, the energy of neutral hadrons is
obtained from the corresponding calibrated ECAL and HCAL energies.

Jet reconstruction is performed using the anti-$k_\textrm{t}$
algorithm~\cite{ak5} with a distance parameter of 0.5 in
$\eta$-$\phi$ space.
Jet energy scale corrections~\cite{jec} derived from data and Monte
Carlo (MC) simulation are applied to account for the non-linear
and non-uniform response of the HCAL.
A small residual correction factor is included to correct for
differences in jet response between the detector and its simulation.
The combined corrections are on the order of 5--10\%,
and their corresponding uncertainties range from 1 to 6\%,
depending on the jet pseudorapidity and $\pt$. Jet quality
criteria~\cite{jetid} are applied to remove misidentified jets,
which arise primarily from calorimeter noise.
For both data and simulated signal events, more than 99.8\%
of all selected jets satisfy these criteria. After imposing all
selection requirements, 114\,599 events remain.

The benchmark signal model for the analysis features pair-produced gluinos,
whose production and decay are simulated using the
{\sc pythia}~\cite{pythia} v6.424 event simulation program with the
CTEQ6L1 parton distribution function (PDF) set~\cite{cteqPDFs}.
The gluino decay is modeled as an effective RPV
coupling between the gluino and three light quarks,
with a branching fraction of 100\%.
All superpartners except the gluino are decoupled,
such that the mass of the squark is assumed to be much larger than
that of the gluino~\cite{rouven}. The natural
width of the gluino resonance is taken to be much smaller than the
resolution of the detector, and no intermediate particles are produced in
the gluino decay. The gluino mass is
varied between 250 and 500\GeV in steps of 50\GeV, and additional
mass points are generated at 750, 1000, 1250, and 1500\GeV.
The leading-order (LO) cross section from \PYTHIA is 92\unit{pb} for
a gluino mass of 250\GeV and 2.1\unit{pb} for a gluino mass of
450\GeV, falling to about $1.3\times10^{-5}$\unit{pb}
for a gluino mass of 1500\GeV.
Next-to-leading-order (NLO) correction factors (K-factors)
are calculated using the \PROSPINO~\cite{prospino} program and are
applied to the LO cross sections. These values rise from 1.8 for
a gluino mass of 250\GeV to 2.1 for a gluino mass of 450\GeV,
and to 4.6 for a gluino mass of 1500\GeV, with corresponding
uncertainties on the NLO cross sections of 15.1, 15.9, and 27.9\%,
respectively. Simulation of the CMS
detector response is performed using \GEANTfour~\cite{geant4}.

\begin{figure}
\centering
\includegraphics[width=\cmsFigWidth]{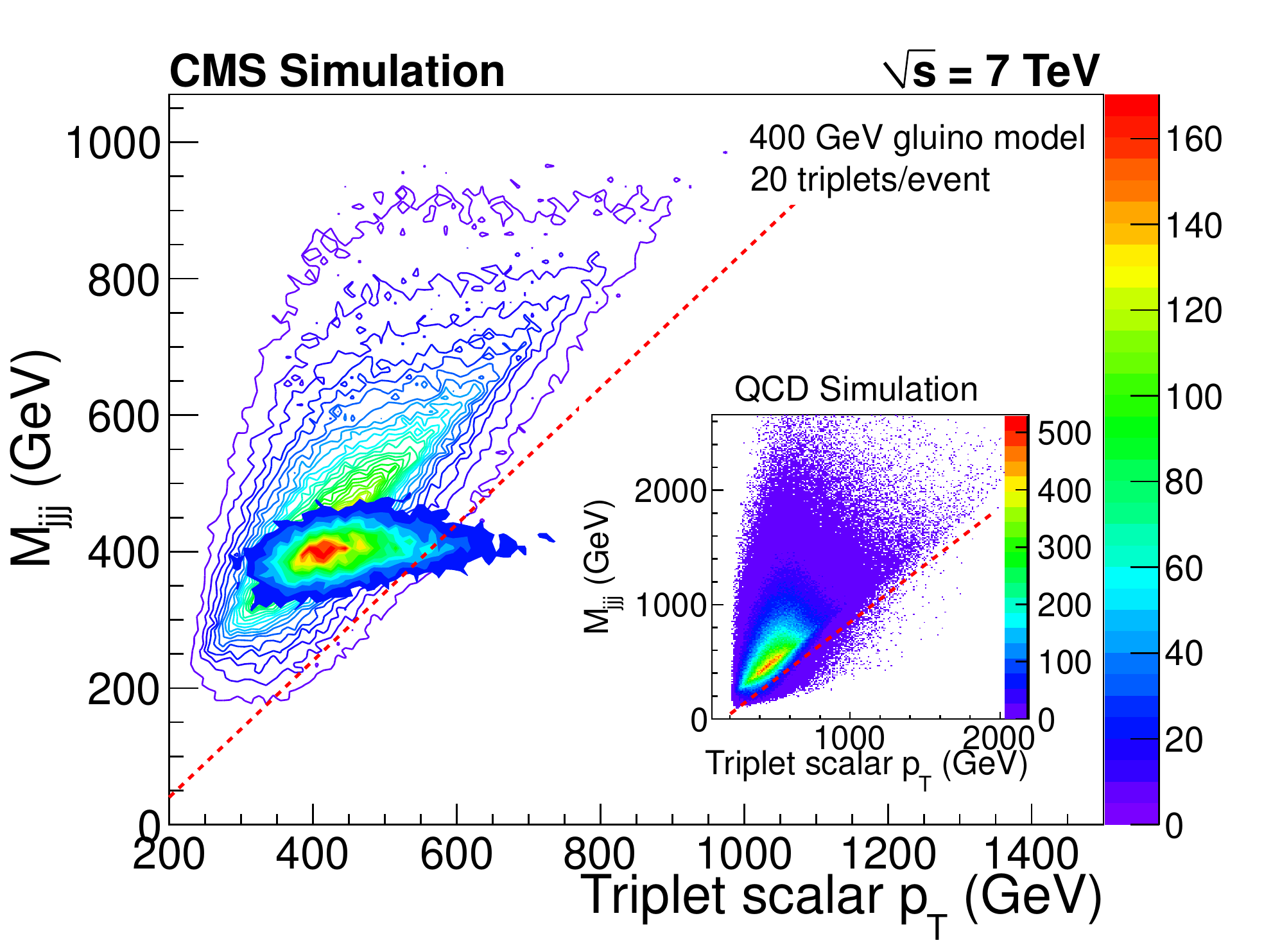}
\caption{Simulated triplet jet mass $M_{\textrm{jjj}}$ versus
         the triplet scalar $\pt$ of all 20
         triplets for a gluino mass of 400\GeV,
         shown as a contoured distribution, with that
	 of the triplets whose jets all originate
	 from the same gluino parent overlaid.
	 The overlaid ``correct'' triplets are shown
	 to lie along a horizontal line corresponding
	 to the generated gluino mass, while the behavior
	 of all other triplets resembles the expectation
	 for the QCD background, which is shown in the
	 inset. All triplets falling to the right of the
	 dashed line pass the requirement of
	 Eq.~(\ref{eq:diag_def}) for a value of $\Delta=160$\GeV.}
\label{fig:2DTrips}
\end{figure}

The principal background arises from multijet events produced in
strong interaction (QCD) processes. All other SM processes, such
as fully-hadronic decays of top quark pairs, are predicted to make
a negligible contribution. The QCD background arises primarily from events
with two high-$\pt$ jets from a hard interaction, combined with
gluon jets from initial- or final-state radiation.
We use the jet-ensemble technique to reconstruct the gluino
candidates and to suppress background. The six highest-$\pt$ jets in
each event are
combined in all possible three-jet combinations (triplets), resulting
in 20 unique triplets per event. For signal events, at most two of the
triplets correspond to a parent gluino, while the other triplets
constitute a combinatorial background. Thus, background arises not only from
QCD multijet events, but also from the signal events themselves.
To improve sensitivity to the presence of a three-jet resonance, a requirement
that exploits the observed linear correlation between the triplet mass and the
scalar sum of the three jets' $\pt$ values in background triplet combinations
is imposed. We require
\begin{eqnarray}
M_{\textrm{jjj}} \,<\,\sum_{i=1}^{3}p_\mathrm{T}^{i}\,-\,\Delta,
\label{eq:diag_def}
\end{eqnarray}
where $M_{\textrm{jjj}}$~is the triplet mass,~$\sum_{i=1}^{3}p_\mathrm{T}^{i}$
is the sum of the magnitudes of the transverse momenta of the
jets within the triplet, and $\Delta$ is an offset adjusted to
optimize signal sensitivity. Note that for signal triplets,
$M_{\textrm{jjj}}$ is fixed by the gluino mass and thus independent of
its three jets' scalar $\pt$ sum. Figure~\ref{fig:2DTrips} shows
the simulated triplet mass versus the triplet scalar $\pt$ for a
gluino mass of 400\GeV.
The value of $\Delta$ is
determined by maximizing the ratio of the number of
signal triplets to the sum of the number of signal and
background triplets in a two-standard-deviation ($\sigma$)
interval about the mean of a Gaussian function that
describes the gluino mass peak. A common value
$\Delta = 160$\GeV, which maximizes sensitivity to signal,
is used for all gluino masses considered, and
typically at most one triplet per event
survives the requirement from Eq.~(\ref{eq:diag_def}).
The fraction of signal triplets, whose jets all originate
from the same parent gluino, that pass this requirement
ranges from 2 to 13\% depending on the gluino mass, while
the efficiency to select all other triplet combinations in
a signal event is below 1\%.

After the final selection, residual background remains from both QCD
multijet events and the combinatorial background triplets in possible gluino
signal events. The latter contribute only minimally, and the shape
of their mass distribution is found to be similar to that of the dominant
QCD multijet background. We therefore consider the two background
components together and parameterize their shape using the
smoothly falling distribution~\cite{dijetmass} given by:
\begin{eqnarray}
\frac{\rd\sigma}{\rd M_{\textrm{jjj}}} = \frac{P_0 (1-M_{\textrm{jjj}}/\sqrt{s})^{P_1}}{(M_{\textrm{jjj}}/\sqrt{s})^{P_2 + P_3 \ln(M_{\textrm{jjj}}/\sqrt{s})}}.
\label{eqn:fourparam}
\end{eqnarray}
\noindent
The parameters $P_0$, $P_1$, $P_2$, and $P_3$ have
values of ($61.0\pm1.8)\times10^{3}\GeV^{-1}$, $53.8\pm0.7$,
$-1.90\pm0.04$, and $-0.13\pm0.01$, respectively, as determined
by fitting this function to the data sample in the range from
260 to 1625\GeV using a $\chi^2$ minimization technique. Thus,
any potential signal would manifest itself as a localized
positive deviation from the expectation of the background fit.

To estimate the number of signal events expected after all selection
criteria are applied, the sum of a Gaussian function that represents
the signal and the four-parameter function (Eq.~(\ref{eqn:fourparam}))
that models the background are fitted to the simulated gluino $M_{\textrm{jjj}}$
distribution, for $M_{\textrm{jjj}}>260$\GeV.
Below 260\GeV, the triplet mass distribution falls off
rapidly due to restrictions imposed by the $\HT$ threshold in the HLT.
The width of the Gaussian function modeling the signal varies according
to the detector resolution from 17\GeV for a gluino mass of
250\GeV, to 100\GeV for a gluino mass of 1500\GeV,
as determined from simulation. The integral of the Gaussian
component provides the estimate for the expected number of signal
triplets reconstructed, and the value of this integral, divided by the
number of signal events generated, determines the signal acceptance
for each gluino mass.  The signal acceptance per event includes all
selection criteria and is parameterized
with a second-order polynomial as a function of gluino mass. The event
acceptance is 0.25\% for a gluino of mass 250\GeV, 1.5\% for a
gluino of mass 450\GeV, and rises to 2.6\% for a gluino of mass
1500\GeV.

The systematic uncertainty on the signal acceptance is evaluated as
follows. The uncertainty related to the jet energy scale~\cite{jec}
is evaluated by varying the jet energy scale correction within its
uncertainties,
then recalculating the acceptance for different gluino mass values.
The largest difference with respect to the nominal
acceptance is taken as the systematic uncertainty and ranges from
9 to 18\%. The difference between the calculated gluino acceptance
at each mass point and the parameterized acceptance is taken as an
uncertainty for each mass value and varies from 1 to 18\%.
The level of initial- and final-state radiation is varied following
a prescription used in many other results~\cite{topmass}, where
the relative amount of each is separately increased and decreased
with respect to the nominal value.
The associated uncertainty is evaluated in a similar manner to that
described for the jet energy scale uncertainties. The difference
of 3 to 5\% with respect to the nominal acceptance is taken as the
systematic uncertainty. To determine the effect of pileup on the
signal acceptance, the MC signal samples are reweighted such that
the distribution of reconstructed primary vertices is shifted high
and then low by one standard deviation compared to
the nominal analysis (in the nominal analysis, the MC distribution
is reweighted to correspond to the measured distribution).
The acceptance procedure is repeated at both points and the largest
difference with respect to the nominal acceptance is taken
as the systematic uncertainty, which varies from 1 to 3\%.
The uncertainty corresponding to the choice of PDF set is evaluated
by reweighting the events to correspond to each of its associated
eigenvector sets and repeating the acceptance calculation. The
difference, added in quadrature from all sets, is 4\% and is
taken as the systematic uncertainty. To estimate the uncertainty
due to the choice of background parameterizations, alternate
functions were fit to the data and compared to the default fit. The
results of each of these fits agree within statistical uncertainties,
and the largest deviation from any of the alternate fits to the default
fit is taken as the
systematic uncertainty, ranging from 1.3\% at a mass of 300\GeV
to 2.9\% at a mass of 500\GeV. These contributions, combined with
those from the luminosity measurement (2.2\%), yield a total systematic
uncertainty on the signal acceptance between 14 and 26\%, depending
on the gluino mass.

\begin{figure}
\centering
\includegraphics[width=\cmsFigWidth]{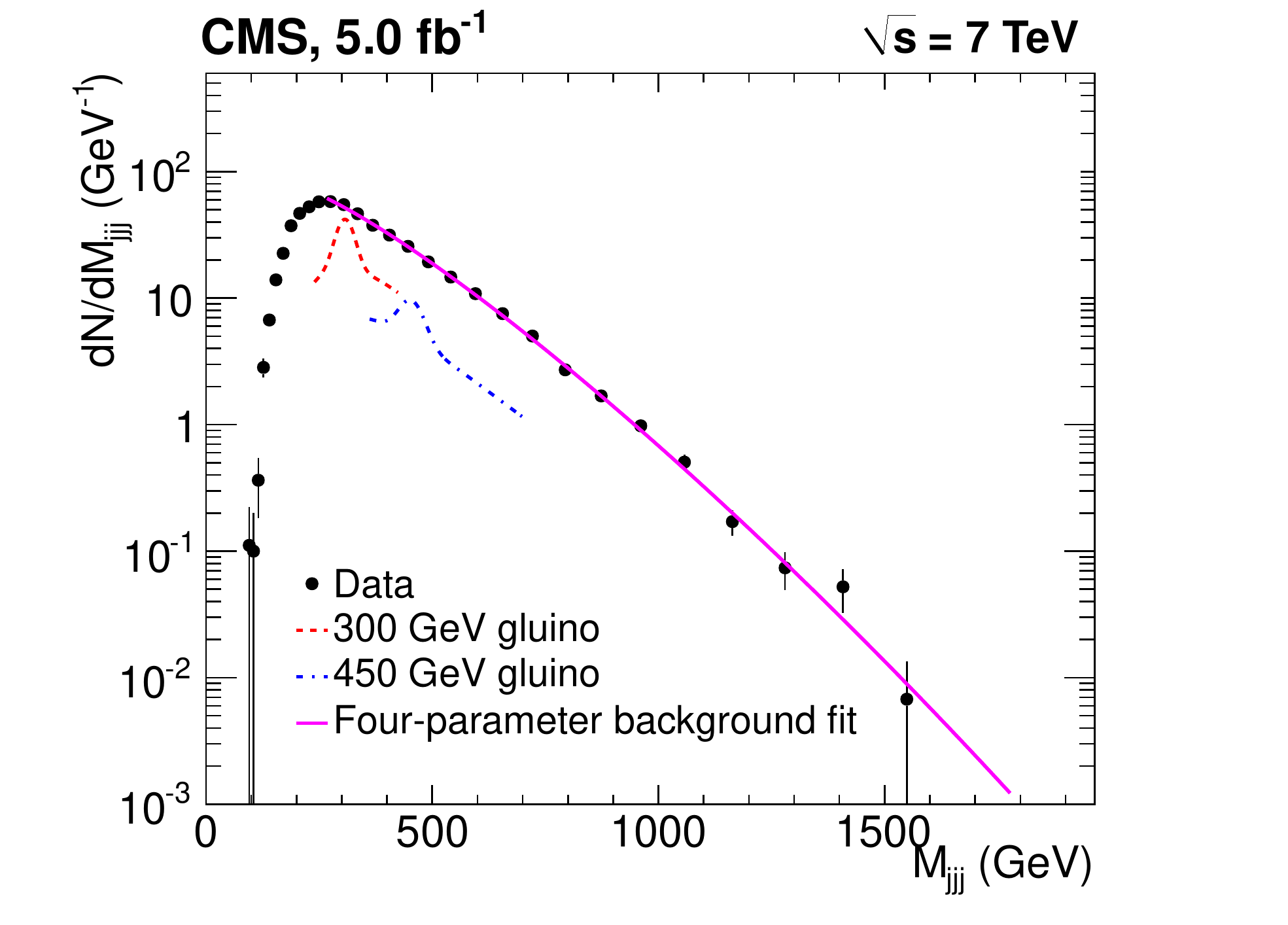}
\caption{
         Mass distribution for selected jet triplets.
         The four-parameter background fit is shown
         by the solid line. The NLO expectation for a
         300\GeV gluino signal is shown by the dotted
         line, and that of a 450\GeV gluino signal by
         the dash-dotted line. Each of the two is
         normalized to the integrated luminosity of
         the data sample.}
\label{fig:DataFitB}
\end{figure}

The measured mass distribution for the selected jet triplets is
shown in Fig.~\ref{fig:DataFitB}, where the four-parameter
background fit is represented by the solid line. The expectation
for a 300\GeV and a 450\GeV gluino signal are also included as
a dotted and dash-dotted line, respectively, and each is
normalized to the integrated luminosity of the data sample.
We choose the width of bins to equal 10\% of the mass value. This
yields bin widths in general correspondence with the expected
instrumental mass resolution.

\begin{figure}
\centering
\includegraphics[width=0.4\textwidth]{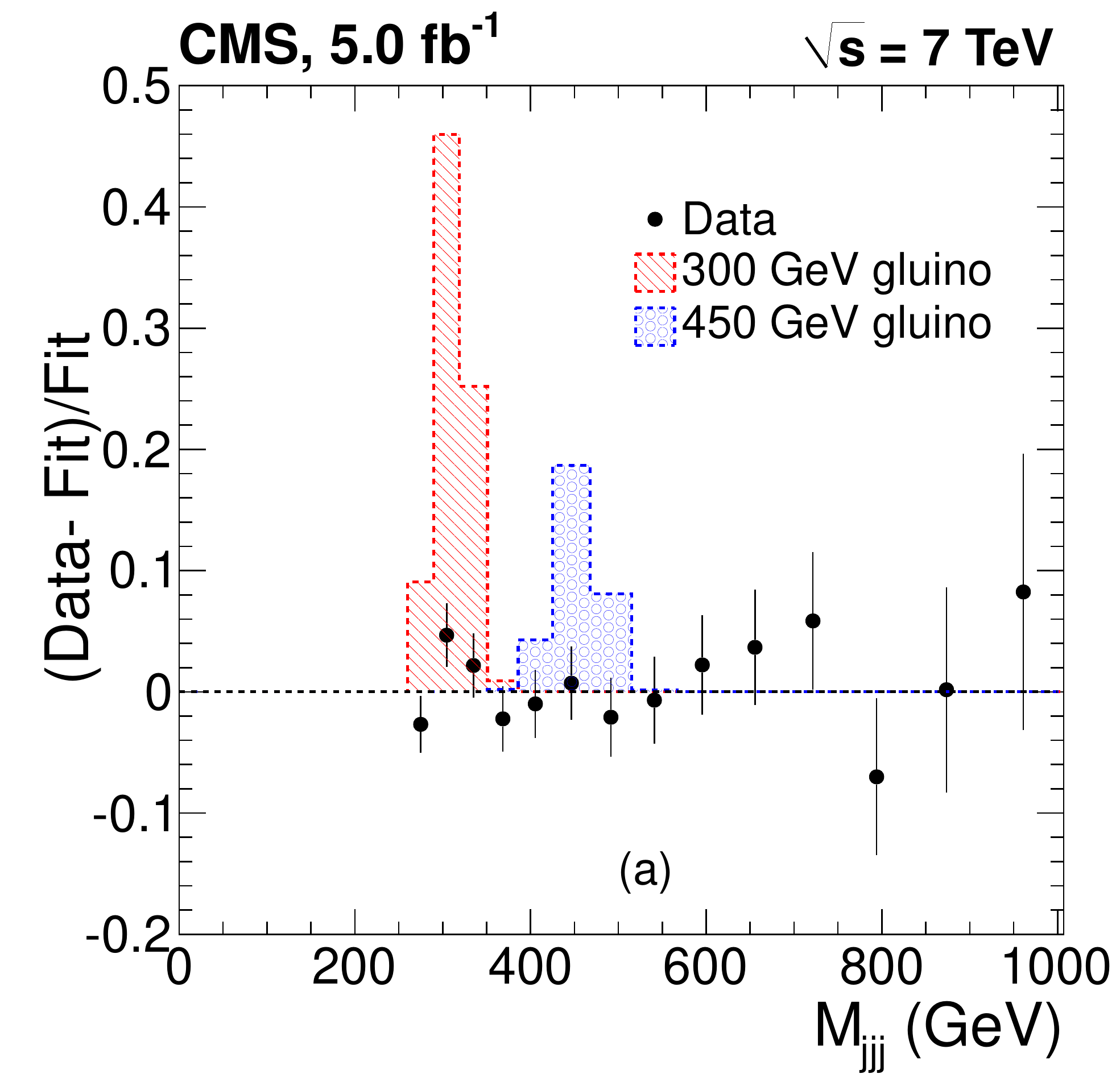}
\includegraphics[width=0.4\textwidth]{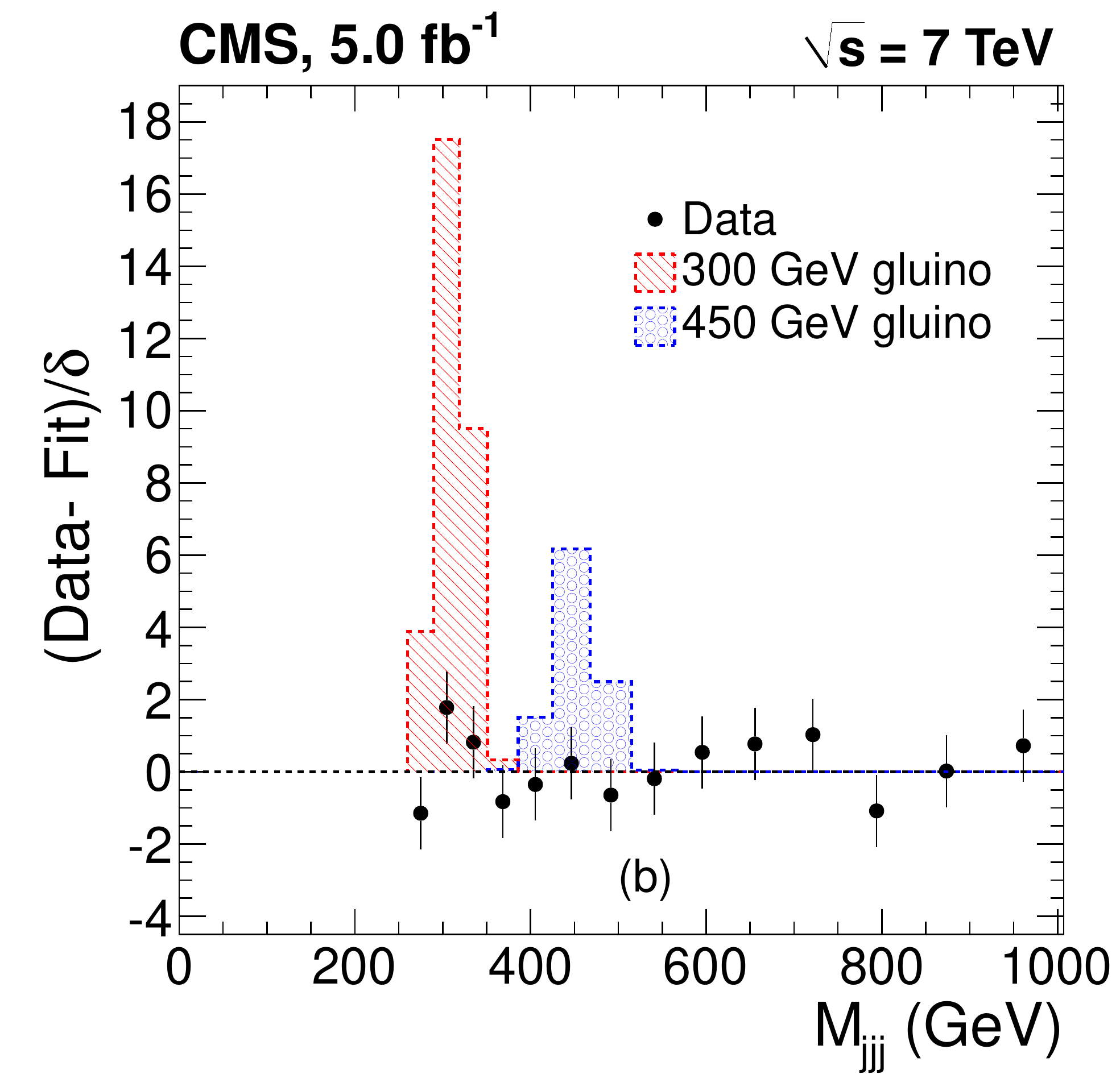}
\caption{Difference between the measured triplet mass
         distribution and the fitted background
         parametrization, divided by the fitted value
         (a) or by the statistical uncertainty $\delta$
         on the fitted value (b), shown for both data
         and the NLO expectation of two gluino models,
         one with a gluino mass of 300\GeV and the
         other with a gluino mass of 450\GeV.}
\label{fig:pullresplot}
\end{figure}

To search for an excess in the number of triplets with respect
to the expected background, we compute the difference between the
measured number of triplets observed in each mass bin and the
number of triplets predicted in this bin by the background fit.
Figure~\ref{fig:pullresplot} shows
this difference, normalized to the background prediction
(Fig.~3a) and to the uncertainty in the background prediction
(Fig.~3b). In both cases, the predicted background is in
agreement with the data over the full mass range.

Because no excess of events is observed in the measured mass
distribution, we proceed to place upper limits on the cross
section times the branching fraction for the production of
three-jet resonances in the data. Both observed and expected
limits are calculated using a modified frequentist CL$_\textrm{s}$ method
with a one-sided profile likelihood test statistic~\cite{higgscombo}.
The background model parameters and their
corresponding uncertainties are taken from the fit of the
four-parameter background function (Eq.~(\ref{eqn:fourparam})) to the
data. The uncertainties on the parameters that describe the
background shape are included as log-normal constraints, and
they are 1.3\% for $P_1$, 2.2\% for $P_2$, 8.6\% for $P_3$, with
a normalization uncertainty on $P_0$ of 3\%. The central value of
each parameter is set to the best fit value and the width to one
standard deviation. The range is truncated at $\pm 3 \sigma$. In
addition to the background parameters, log-normal constraints are included
for the acceptance and integrated luminosity.
These systematic uncertainties are treated as global constraints in the likelihood.

\begin{figure}
\centering
\includegraphics[width=\cmsFigWidth]{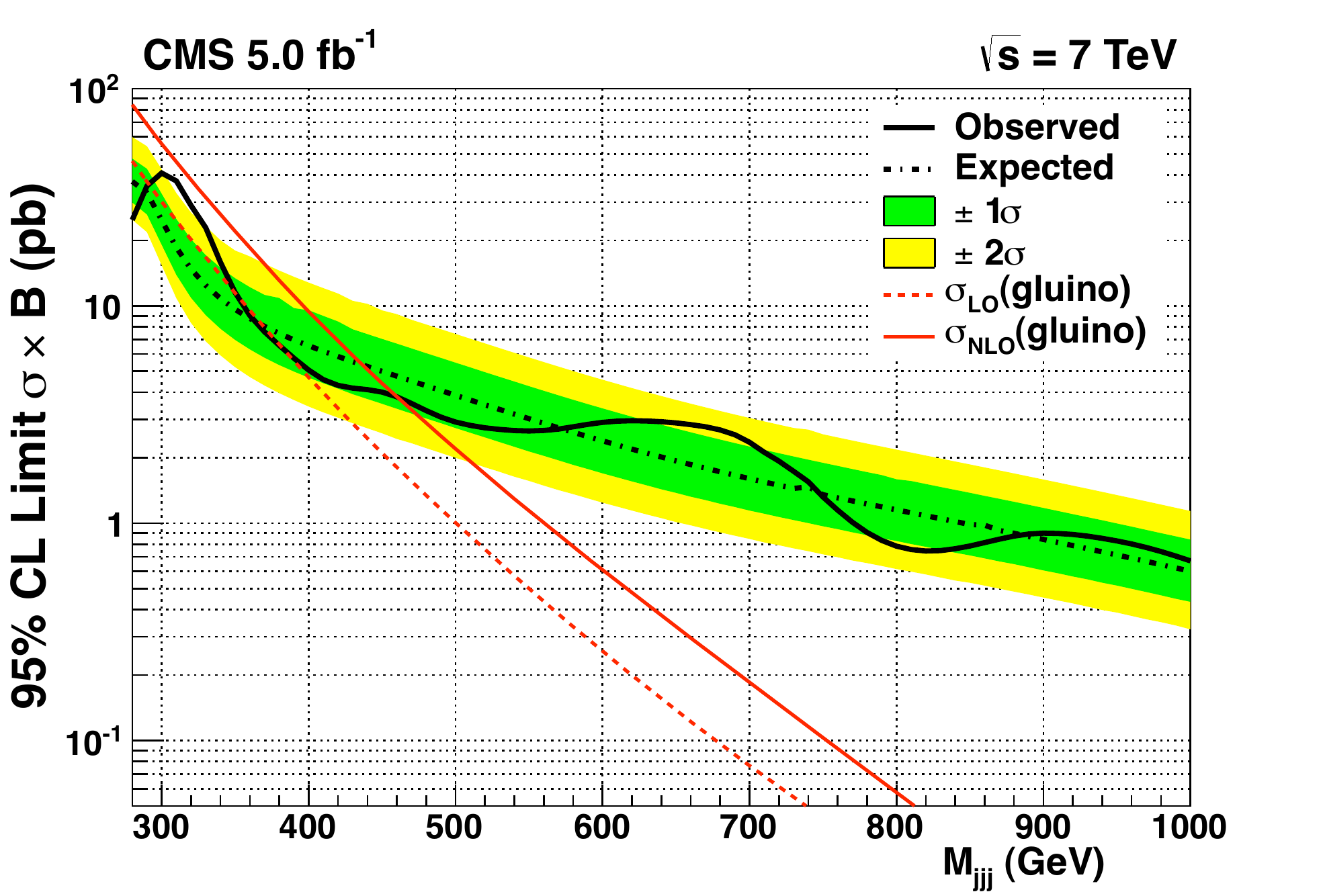}
\caption{Observed and expected 95\% CL upper limits
on the cross section times branching fraction for
gluino pair production followed by RPV decay of each
gluino to three light-flavored quark jets. Also shown
are the $\pm1\sigma$ and $\pm2\sigma$ bands on the
expected limit, as well as the theoretical LO and NLO
cross sections for gluino production, assuming a
branching fraction of a gluino decay into three jets
of 100\%.}
\label{fig:Limits}
\end{figure}

The observed and expected 95\% confidence level (CL) upper limits on the
gluino pair production cross section times branching fraction as a function
of gluino mass are presented in Fig.~\ref{fig:Limits}.
The corresponding 95\% CL lower limit on the gluino mass is determined by
finding the mass value at which the limit line crosses that of the
NLO gluino cross section, assuming a branching fraction for the signal
model of 100\%. We perform the search in the region of $M_{\textrm{jjj}}$ above
280\GeV and exclude masses below 460\GeV at 95\% CL.

In summary, a search has been performed for three-jet resonance production in
$\Pp\Pp$ collisions at a center-of-mass energy of 7 TeV, using a data sample
corresponding to an integrated luminosity of 5.0\fbinv. Events having the
properties of high jet multiplicity and large scalar sum of jet $\pt$, which
are expected signatures of high-mass hadronic resonances, were analyzed for
the presence of signal events, and no evidence for a narrow resonance was found.
For a mass range above 280\GeV, the production of gluinos,
modeled as an effective RPV coupling between the gluino and three
light quarks with a branching fraction of 100\%, has been
excluded for masses below 460\GeV at 95\% CL. These are the most stringent
limits to date.

The authors would like to thank Michael Park and Yue Zhao for providing
theoretical calculations.
We congratulate our colleagues in the CERN accelerator departments for
the excellent performance of the LHC machine. We thank the technical
and administrative staff at CERN and other CMS institutes, and acknowledge
support from BMWF and FWF (Austria); FNRS and FWO (Belgium); CNPq,
CAPES, FAPERJ, and FAPESP (Brazil); MES (Bulgaria); CERN; CAS, MoST,
and NSFC (China); COLCIENCIAS (Colombia); MSES (Croatia); RPF (Cyprus);
MEYS (Czech Republic); MoER, SF0690030s09 and ERDF (Estonia); Academy of
Finland, MEC, and HIP (Finland); CEA and CNRS/IN2P3 (France); BMBF, DFG,
and HGF (Germany); GSRT (Greece); OTKA and NKTH (Hungary); DAE and DST
(India); IPM (Iran); SFI (Ireland); INFN (Italy); NRF and WCU (Korea);
LAS (Lithuania); CINVESTAV, CONACYT, SEP, and UASLP-FAI (Mexico); MSI
(New Zealand); PAEC (Pakistan); MSHE and NSC (Poland); FCT (Portugal);
JINR (Armenia, Belarus, Georgia, Ukraine, Uzbekistan); MON, RosAtom, RAS
and RFBR (Russia); MSTD (Serbia); SEIDI and CPAN (Spain); Swiss Funding
Agencies (Switzerland); NSC (Taipei); TUBITAK and TAEK (Turkey); NASU
(Ukraine); STFC (United Kingdom); DOE and NSF (USA). Individuals have
received support from the Marie-Curie programme and the European
Research Council (European Union); the Leventis Foundation; the A. P.
Sloan Foundation; the Alexander von Humboldt Foundation; the Austrian
Science Fund (FWF); the Belgian Federal Science Policy Office; the
Fonds pour la Formation \`a la Recherche dans l'Industrie et dans
l'Agriculture (FRIA-Belgium); the Agentschap voor Innovatie door
Wetenschap en Technologie (IWT-Belgium); the Council of Science and
Industrial Research, India; the Compagnia di San Paolo (Torino); and
the HOMING PLUS programme of Foundation for Polish Science, cofinanced
from European Union, Regional Development Fund.

\bibliography{auto_generated}   

\cleardoublepage \appendix\section{The CMS Collaboration \label{app:collab}}\begin{sloppypar}\hyphenpenalty=5000\widowpenalty=500\clubpenalty=5000\textbf{Yerevan Physics Institute,  Yerevan,  Armenia}\\*[0pt]
S.~Chatrchyan, V.~Khachatryan, A.M.~Sirunyan, A.~Tumasyan
\vskip\cmsinstskip
\textbf{Institut f\"{u}r Hochenergiephysik der OeAW,  Wien,  Austria}\\*[0pt]
W.~Adam, E.~Aguilo, T.~Bergauer, M.~Dragicevic, J.~Er\"{o}, C.~Fabjan\cmsAuthorMark{1}, M.~Friedl, R.~Fr\"{u}hwirth\cmsAuthorMark{1}, V.M.~Ghete, J.~Hammer, N.~H\"{o}rmann, J.~Hrubec, M.~Jeitler\cmsAuthorMark{1}, W.~Kiesenhofer, V.~Kn\"{u}nz, M.~Krammer\cmsAuthorMark{1}, I.~Kr\"{a}tschmer, D.~Liko, I.~Mikulec, M.~Pernicka$^{\textrm{\dag}}$, B.~Rahbaran, C.~Rohringer, H.~Rohringer, R.~Sch\"{o}fbeck, J.~Strauss, A.~Taurok, W.~Waltenberger, G.~Walzel, E.~Widl, C.-E.~Wulz\cmsAuthorMark{1}
% ...
\vskip\cmsinstskip
\dag:~Deceased\\
1:~~Also at Vienna University of Technology, Vienna, Austria\\
% ...

\end{sloppypar}
\end{document}